\documentclass[a4paper,prb,superscriptaddress,citeautoscript,twocolumn,twoside]{revtex4}

\usepackage{color}
\usepackage{graphicx}
\usepackage[utf8x]{inputenc}
\usepackage[T1]{fontenc}
\usepackage{siunitx}
\usepackage{amstext}
\usepackage{textgreek}
\usepackage{hyperref}

\usepackage{epstopdf}
\usepackage{amsfonts}
\usepackage{amsmath}
\usepackage{mathtools}
\usepackage{multirow}
\usepackage{dsfont}
\graphicspath{{../},{figures/}}
\usepackage{float}

\makeatletter

\newcommand{\Rnum}[1]{\expandafter\@slowromancap\romannumeral #1@}

\begin{document}

\title{Experimentally attacking quantum money schemes based on quantum retrieval games}

\author{Kateřina Jiráková}
\email{katerina.jirakova@upol.cz}
\affiliation{RCPTM, Joint Laboratory of Optics of Palacký University and Institute of Physics of Czech Academy of Sciences, 17. listopadu 12, 771 46 Olomouc, Czech Republic}
\thanks{Presently on leave at Faculty of Physics, Adam Mickiewicz University,
PL-61-614 Pozna\'n, Poland}

\author{Karol Bartkiewicz} \email{bark@amu.edu.pl}
\affiliation{RCPTM, Joint Laboratory of Optics of Palacký University and Institute of Physics of Czech Academy of Sciences, 17. listopadu 12, 771 46 Olomouc, Czech Republic}
\affiliation{Faculty of Physics, Adam Mickiewicz University,
PL-61-614 Pozna\'n, Poland}

\author{Antonín Černoch} \email{acernoch@fzu.cz}
\affiliation{Institute of Physics of the Czech Academy of Sciences, Joint Laboratory of Optics of PU and IP AS CR, 17. listopadu 50A, 772 07 Olomouc, Czech Republic}
   
\author{Karel Lemr}
\email{k.lemr@upol.cz}
\affiliation{RCPTM, Joint Laboratory of Optics of Palacký University and Institute of Physics of Czech Academy of Sciences, 17. listopadu 12, 771 46 Olomouc, Czech Republic}   
\thanks{Presently on leave at  Faculty of Physics, Adam Mickiewicz University,
PL-61-614 Pozna\'n, Poland}

\begin{abstract}
The concept of quantum money (QM) was proposed by Wiesner in the 1970s. Its main advantage is that every attempt to copy QM  unavoidably leads to imperfect counterfeits. In the Wiesner's protocol, quantum banknotes need to be delivered to the issuing bank for verification. Thus, QM requires quantum communication which range is limited by noise and losses. Recently, Bozzio et al. (2018) have demonstrated experimentally how to replace challenging quantum verification with a classical channel and a quantum retrieval game (QRG). This brings QM significantly closer to practical realisation, but still thorough analysis of the revised scheme QM is required before it can be considered secure. We address this problem by presenting a proof-of-concept attack on QRG-based QM schemes, where we show that even imperfect quantum cloning can, under some circumstances, provide enough information to break a QRG-based QM scheme. 
\end{abstract}

\maketitle
\section*{Introduction}

All payment methods are potential targets of thieves and counterfeiters. Over the course of history, we have witnessed a race of arms between the counterfeiters and issuers of various currencies. Remarkably, Sir Isaac Newton, who became the master of Royal Mint, enforced laws against counterfeiting. Nevertheless, the methods used by Newton become obsolete when it comes to modern payment methods. With the rapid technological progress, we are beginning to consider a situation where counterfeiting is no longer limited by the available technology, but rather by the laws of nature. An example of such fundamental limitation is the no-cloning theorem,\cite{Wootters1982,Dieks1982} which guaranties security of quantum money \cite{Wiesner1983,LemrCernoch2017}.

In a recent paper, Bozzio \emph{et al.} \cite{Bozzio2018} reported on an implementaion of a QM scheme based on QRGs \cite{Amiri2017,Bar-Yossef2004}. While this result brings QM closer to practical implementation, here we demonstrate that QRG-based QM schemes are still vulnerable to a new kind of attack (for some typical attacks see Ref.~[\!\!\citenum{Bartkiewicz2013,Gisin2002,Bechmann1999}]) which can be considered a quantum version of sniffing (a hacking method used to monitor clasical information). The general idea of our attack can be used against a broader range of QM schemes based on QRG~\cite{Gavinsky2012, Pastawski2012, Georgiou2015} and potentially on other quantum communication protocols. 
Thus, our results can facilitate future practical implementations of QM by providing a method for exploring the security limits allowed in QRG-based protocols.
For the purpose of our research we have experimentally recreated the original scheme of Ref.~[\!\!\citenum{Bozzio2018}]. Its working principle can be described as follows: the bank encodes QM (as a quantum token) using a secret sequence of qubit pairs chosen from the list of eight options:
\begin{equation}
S = \{|0+ \rangle, |0- \rangle, |1+ \rangle, |1- \rangle, |+0 \rangle, |-0 \rangle, |+1 \rangle,|-1 \rangle\}\, ,\label{eq:seq}
\end{equation}
where $|0\rangle$, $|1\rangle$ are logical qubit states, and $|\pm\rangle = \frac{1}{\sqrt{2}}\left(|0\rangle\pm|1\rangle\right)$ stand for their superpositions. The sequence and its serial number is stored on a quantum credit card~\cite{Bozzio2018,Wolters2017,Wang2015} subsequently given to a client of the bank. Upon payment, the credit card is inserted into the vendor's terminal which is supposed to perform projection measurements on these pairs in a measurement basis requested by the bank (randomly chosen to be either 0/1 or +/- for an entire pair). Then, the terminal sends the classical outcomes of those measurements to the bank. The main advantage of this scheme is that the terminal measurement itself is sufficient for authentication of the credit card, so quantum states do not have to be sent to the bank for verification. The bank just checks the results knowing the specific encoded states and either accepts or denies the payment. A small amount of errors is expected to appear in the verification procedure to account for implementation imperfections. The acceptable amount of errors needs to be small enough to ensure that payment by a cloned quantum credit card is denied. In contrast to the original Wiesner QM scheme~\cite{Wiesner1983}, no on-line quantum channel has to be used for payment. Thus, the verifiability problem as defined by Aaronson and Christiano~\cite{Aaronson2012} is at least partially solved.

This protocol is secure against a dishonest terminal only if each quantum sequence is generated using a truly random encoding. However, such condition would give rise to a giant database problem, as discussed in~[\!\!\citenum{Aaronson2012}] and~[\!\!\citenum{Bennett1982}]. The random sequence approach is highly impractical or even infeasible.  In practice, there has to be one secret encoding function shared by a certain number quantum banknotes or tokens (i.e., sequences of quantum states and their serial numbers). Hence, in our research we test limitations of sharing a secret encoding by multiple tokens.

\begin{figure}
		\begin{center}
		\includegraphics[width=7cm]{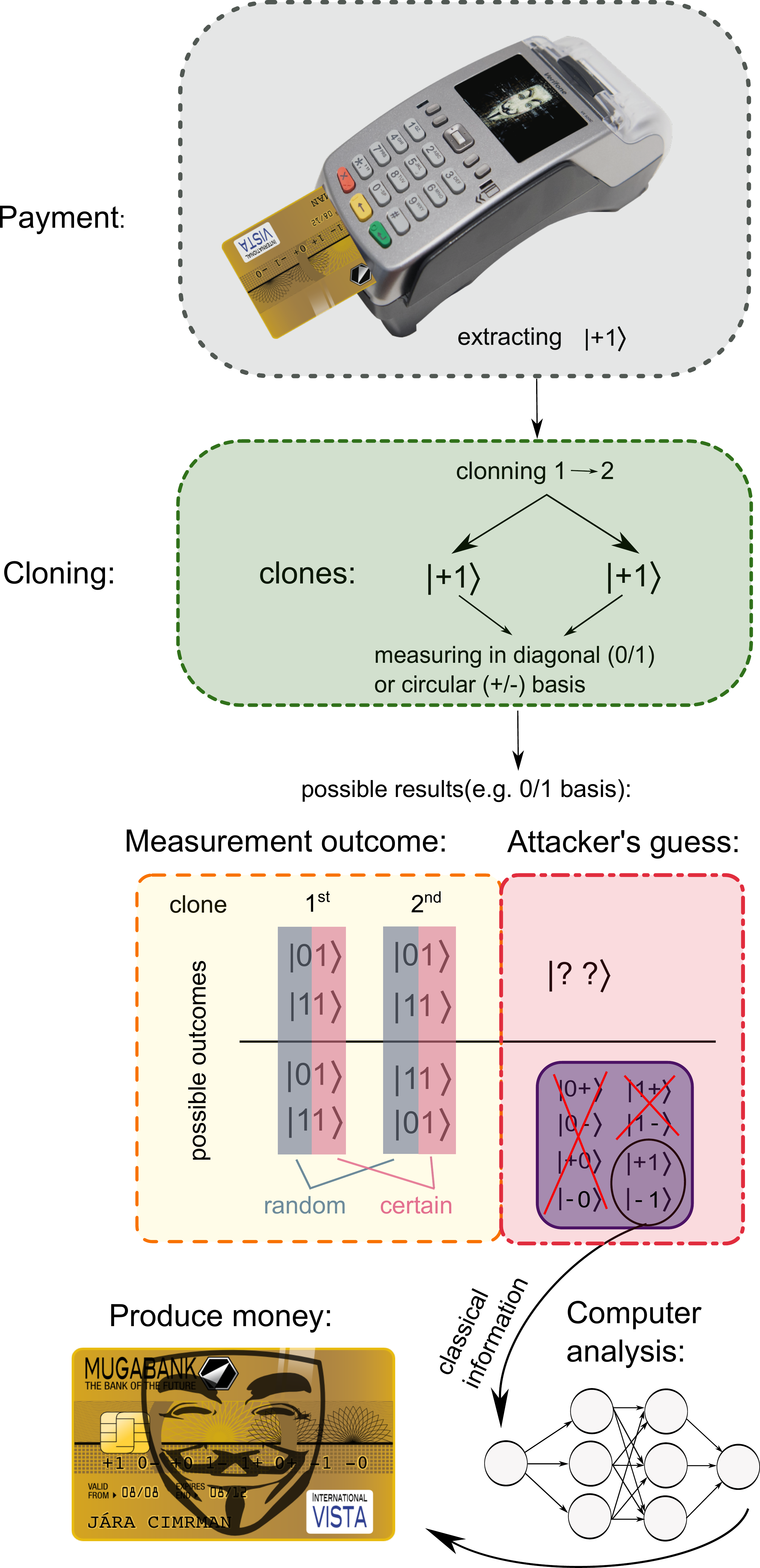}
		\caption{\label{fig:schemeTwo} Attack on a quantum credit card utilising a hacked terminal. During a transaction a pair of states (e.g., $|+ 1\rangle$) is extracted from the card and cloned. Here, for simplicity, we depict only the situation where all the qubits are perfectly copied (the probability of such event is proportional to $F^2$). Then, measurements are performed on all four copies in the basis randomly chosen by the bank (e.g. 0/1). If the measurements on copied qubit pairs produces one of two results from the bottom block of the  table of outcomes, the attacker learns the originally encoded state (in this case $|?\, 1\rangle$). This procedure is repeated  until a relation between the quantum states and serial numbers is learned.  Since then, the attacker can issue perfectly counterfeit quantum credit cards.}	
		\end{center}
\end{figure}

Here, similarly like in Ref.~[\!\!\citenum{LemrCernoch2017}], we use quantum cloning to counterfeit QM. Our approach is virtually undetectable by the bank because we copy only parts of quantum tokens (i.e., quantum sequences). In terms of QRG-based QM protocol, the attacker utilises a compromised payment terminal enabling quantum cloning of an input qubit (see Fig.~\ref{fig:schemeTwo}). The terminal performs measurements on both copies of a qubit providing the attacker with some information on the encoding used by the bank, if two consecutive qubits from a sequence are cloned. The frequency of cloning can be arbitrarily small and therefore made unrecognisable from noise. After gathering enough data, the attacker reveals the secret encoding used by the bank for preparing credit cards. Since then, they can  issue fake quantum credit cards indistinguishable from the original ones issued by the bank.

\section*{Results}

  \begin{figure}
		\begin{center}
		\includegraphics[width=8cm]{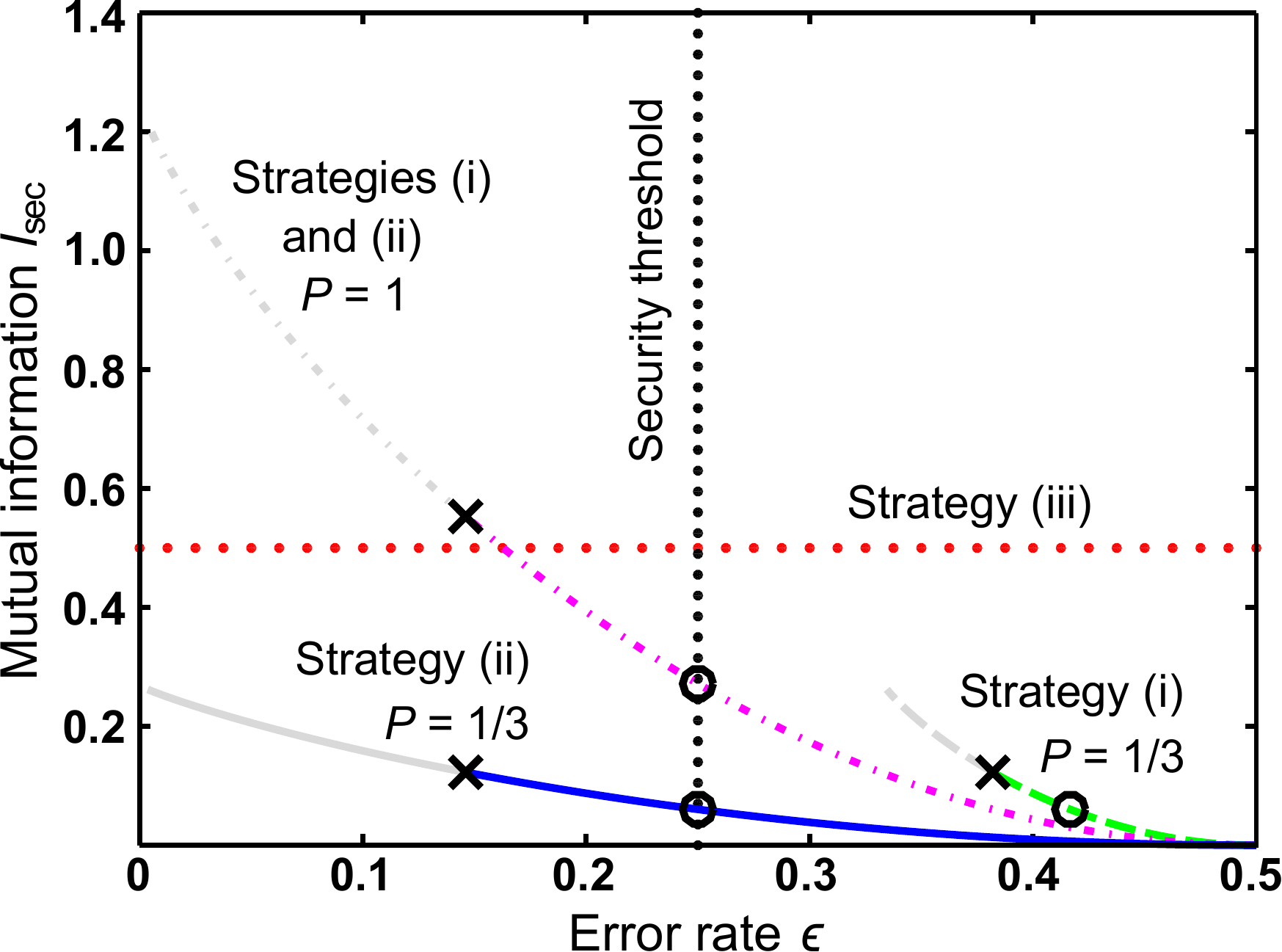}
		\caption{\label{fig:graphMut} Mutual information $I_{\text{sec}}$ versus error rate $\epsilon$ for two fixed probabilities $P = \left\lbrace\frac{1}{3}; 1 \right\rbrace$. Vertical black dotted line represents error rate associated with security threshold discussed in Ref.~[\!\!\citenum{Gavinsky2012}] and~[\!\!\citenum{Pastawski2012}]. Crosses mark the smallest average error introduced by optimal cloning for a fixed value of $P$. Error rates below these optimal values cannot be reached by any physical operation (greyed curves). Circles stand for limit of classical copying ($F = 0.75$). Thus, the segments of curves between circles and crosses mark the regime of quantum copying. It follows from Eq.~\ref{eq:per2} that classical copying limit in strategy (ii) always corresponds to intersection between the relevant curve and the security threshold.}	
		\end{center}
\end{figure}

We have implemented the quantum sniffing attack on the platform of linear optics, where qubits are encoded as polarisation states of single photons. The optimal cloning strategy (i.e., maximizing single-copy cloning fidelity) for copying qubits from the set $S$ is the symmetric phase-covariant cloning (SPCC)~\cite{Bruss2000,Bartkiewicz2013,LemrCernoch2017}. 
In the experiment, pairs of input qubits $|\psi_{1}\psi_{2}\rangle_{\text{in}} \in S$ were subjected to SPCC procedure obtaining two clones  $\hat{\varrho}_{\text{1A}}\otimes\hat{\varrho}_{\text{2A}}$ and $\hat{\varrho}_{\text{1B}}\otimes\hat{\varrho}_{\text{2B}}$ of the input qubit pair. These clones were then measured in the same but random basis. In a QRG-based QM protocol the basis is selected by the bank. Due to limitations of linear optics based implementations of  quantum cloners \cite{Fiurasek2003}, the SPCC process is probabilistic and sometimes it fails to deliver the clones. The probability of successful cloning of one input qubit is denoted $P$. Therefore the probability of cloning the entire qubit pair is $P^2$. Quality of the clones is expressed in terms of fidelity $F$ defined as $F = F_{ij} ={}_{\text{in}}\langle \psi_{i}|\hat{\varrho}_{ij}|\psi_{i}\rangle_{\text{in}},$ where $i = 1, 2$ and $j = \text{A, B}$ denote the first and the second clone,  respectively. The probability of finding both clones $\hat{\varrho}_{i\text{A}}$ and $\hat{\varrho}_{i\text{B}}$ in a given state $|\psi_{i}\rangle_{\text{in}}$ reads $F^2$. An example of an attack on a particular qubit pair is shown in Fig.~\ref{fig:schemeTwo}. 

The theoretical limit for SPCC fidelity~\cite{Bruss2000} is $F = \frac{1}{2} \left(1+\frac{1}{\sqrt{2}}\right) \approx 0.854$ and on the platform of linear optics the cloning succeeds with probability $P = \frac{1}{3}$. While the limit on fidelity is fundamental in its nature, $P$ depends on the physical platform used in a given implementation and can be arbitrarily close to 1. However, even on the platform of linear optics, it is possible to clone at arbitrarily high values of $P$ but at the expense of reaching lower than optimal fidelity $F$ (see hybrid quantum cloners~\cite{Bartkiewicz2013,Bartkiewicz2014}).

The terminal registers two measurement outcomes per input qubit corresponding to the clones. If the two clones of one input qubit yield identical results, while for the other yield opposite results, the attacker gains information about the encoding. With the probability $P_{\textrm{tot}} = P_{\text{c}} + P_{\text{e}}$ the attacker eliminates six of the original eight encodings (see Eq.~\ref{eq:seq}). One of the two remaining encodings have actually been used by the bank. The probability of obtaining correct information from the attack is $P_{\text{c}} = \frac{1}{2}P^2 F^2$, whereas $P_{\text{e}} = \frac{1}{2} P^2(1-F)^2 + P^2 F(1-F)$ stands for the probability of getting an erroneous result due to limited cloning fidelity. Similarly, if the two clones of each input qubit yield identical results, the attacker knows that only one of four encodings might have been sent by the bank.

The attacker is able to learn the method of encoding tokens by accumulating measurement results provided that the fidelity is $F \not=\frac{1}{2}$.
The cloning operation inherently introduces errors in the measurement outcomes~\cite{Wootters1982,Dieks1982}. Hence, the terminal might send to the bank incorrect results. If the error rate surpasses a given limit (25\% in Ref.~[\!\!\citenum{Bozzio2018}]), the bank will reject the payment. Thus, it is necessary to introduce a strategy of attack considering all circumstances of the measurement (i.e., if cloning failed or not) and its outcomes to minimise the error rate. There are generally three distinct strategies: (i) to provide the bank with measurement outcome every time cloning takes place and even if it fails, send a random value, (ii) to send measurement outcome, only if it is registered by the terminal and report a lost qubit when cloning fails and (iii) to measure qubits after their extraction from the credit card in given measurement basis but do not perform cloning at all. 

To quantify the correlations between the attacker and the genuine token we use mutual information $I_{\text{sec}}$, which expresses how many bits of information can the attacker obtain upon cloning one qubit pair. The exact value of mutual information depends on the strategy used, cloning success probability $P$ and fidelity $F$. In case of the third strategy (without cloning), its value is $\tfrac{1}{2}$.

Simultaneously, we denote $\epsilon$ the probability of an error  being reported to the bank. 
The expressions for error rates  $\epsilon$ for the two above-mentioned strategies can be obtained by direct calculations based on analysis of probabilities of all possible scenarios and read
\begin{eqnarray}
\epsilon_{\textrm{(i)}} &=&\tfrac{1}{2} (1-P) + P (1-F)\,  ,\label{eq:per1} \\ 
\epsilon_{\textrm{(ii)}} &=& 1-F\, .\label{eq:per2}
\end{eqnarray}
Equation~(\ref{eq:per1}) takes into account two situations. In the first case, one or both qubits are lost during cloning and, therefore, random results are reported to the bank (50\% chance of error). In the second case, even if cloning succeeds, non-unit fidelity may cause the measurement to yield an incorrect result. The error rate in case of strategy (ii) depends only on imperfect cloning fidelity.

\begin{figure}[H]
		\begin{center}
		\includegraphics[width=8cm]{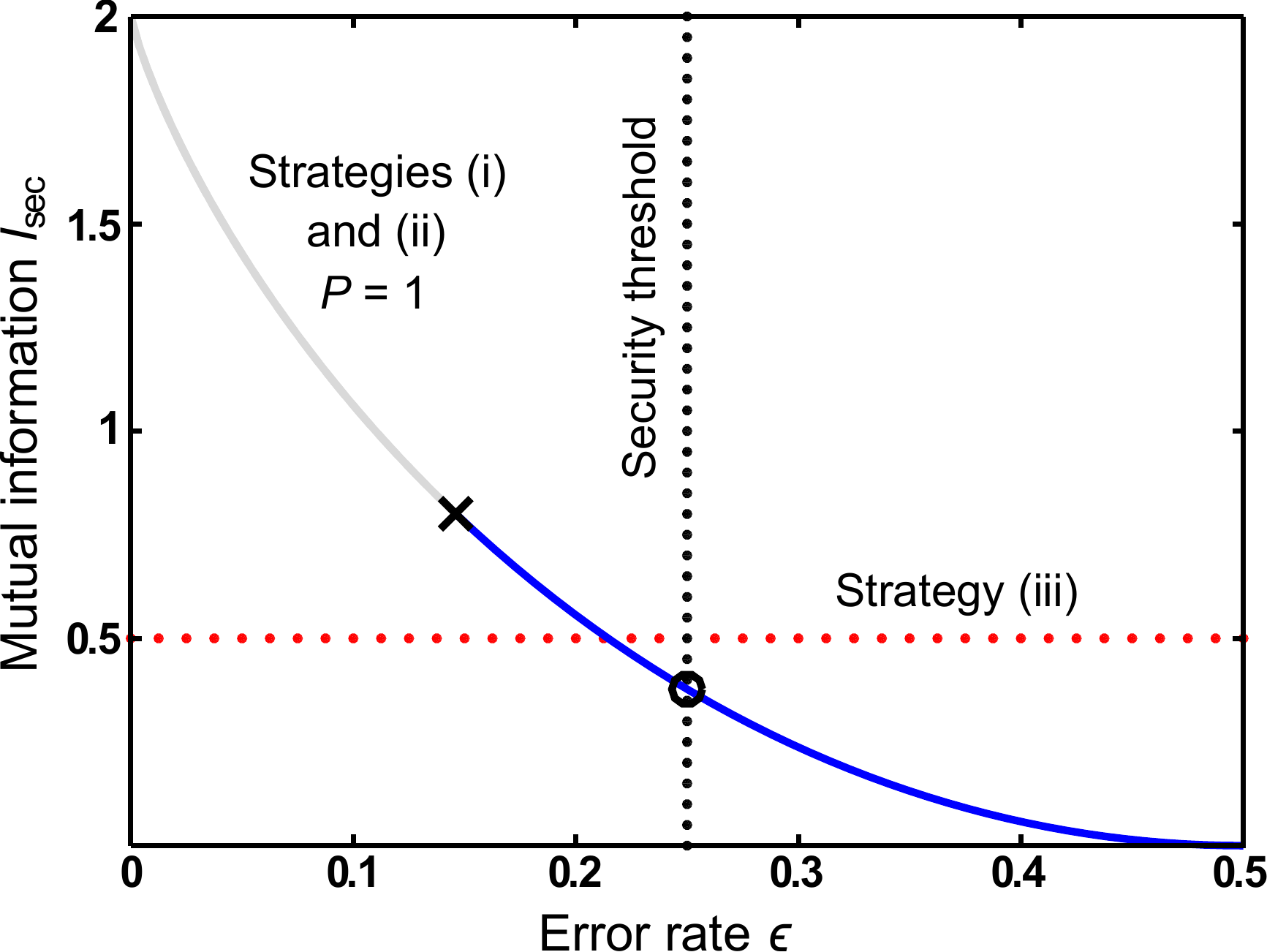}
		\caption{\label{fig:graphMutCond} Conditional mutual information $I_{\text{sec}}$ versus error rate $\epsilon$. Strategies (i) and (ii) are equal in this case.}	
		\end{center}
\end{figure}

The relation between mutual information $I_{\text{sec}}$ (between the bank and the attacker) and the error rate $\epsilon$ for all strategies is show in  Fig.~\ref{fig:graphMut}. In the figure, quantities $I_{\text{sec}}$ and $\epsilon$ are functions of cloning fidelity for $\frac{1}{2}\le F \le 1$ for two cloning success rates $P = \frac{1}{3}$ (linear optics limit~\cite{Fiurasek2003,Bartkiewicz2014,LemrCernoch2017}) and $P = 1$ (deterministic cloning~\cite{Zhang2000,Chefles1999,Bartkiewicz2014,LemrCernoch2017}). In case of deterministic cloning the two attack strategies coincide, but for probabilistic cloning the second strategy provides better results. It is fair to note that the mutual information of any simple linear-optical cloning strategy is lower in comparison with the no-cloning strategy (iii). On the other hand, with deterministic cloning, one can reach even higher values of mutual information and therefore cloning strategies need to be considered for security implications. Additionally, machine learning-based algorithms may require data with as little noise as possible even at the expense of the overall quantity. Post-selection on successful cloning events allows to distil such sample. Corresponding conditional mutual information yields a significantly higher value when both qubits are successfully cloned than for the no-cloning strategy (iii) (Fig. \ref{fig:graphMutCond}).
 
To prove the working principle of the quantum sniffing attack, let us consider a specific encoding of the quantum tokens and demonstrate the attacker's approach to learning the encoding. Here, we assume that the bank uses a hash function to encode the tokens. Hash functions are designed to return very distinct results even for similar inputs. The input can be additionally modified by using a specific secret number (salt). In this case the hash function is often referred to as salted. For simplicity, let us now assume that the hash function is known to the attacker, but the salt is secret. For each token passing through the terminal, the attacker calculates hashes (outputs of the hash function) of its serial number salted by numbers from a certain range. This way the attacker investigates various encodings each corresponding to one secret number (or salt). Using the information gained by quantum sniffing, the attacker calculates the number of agreements (matching qubit pairs) between the predictions of the tested encoding and the measurement outcomes on real tokens. The encoding with highest number of agreements is most probably the one used by the bank, hence the one corresponding to the correct salt.

\begin{figure}
		\begin{center}
		\includegraphics[width=8cm]{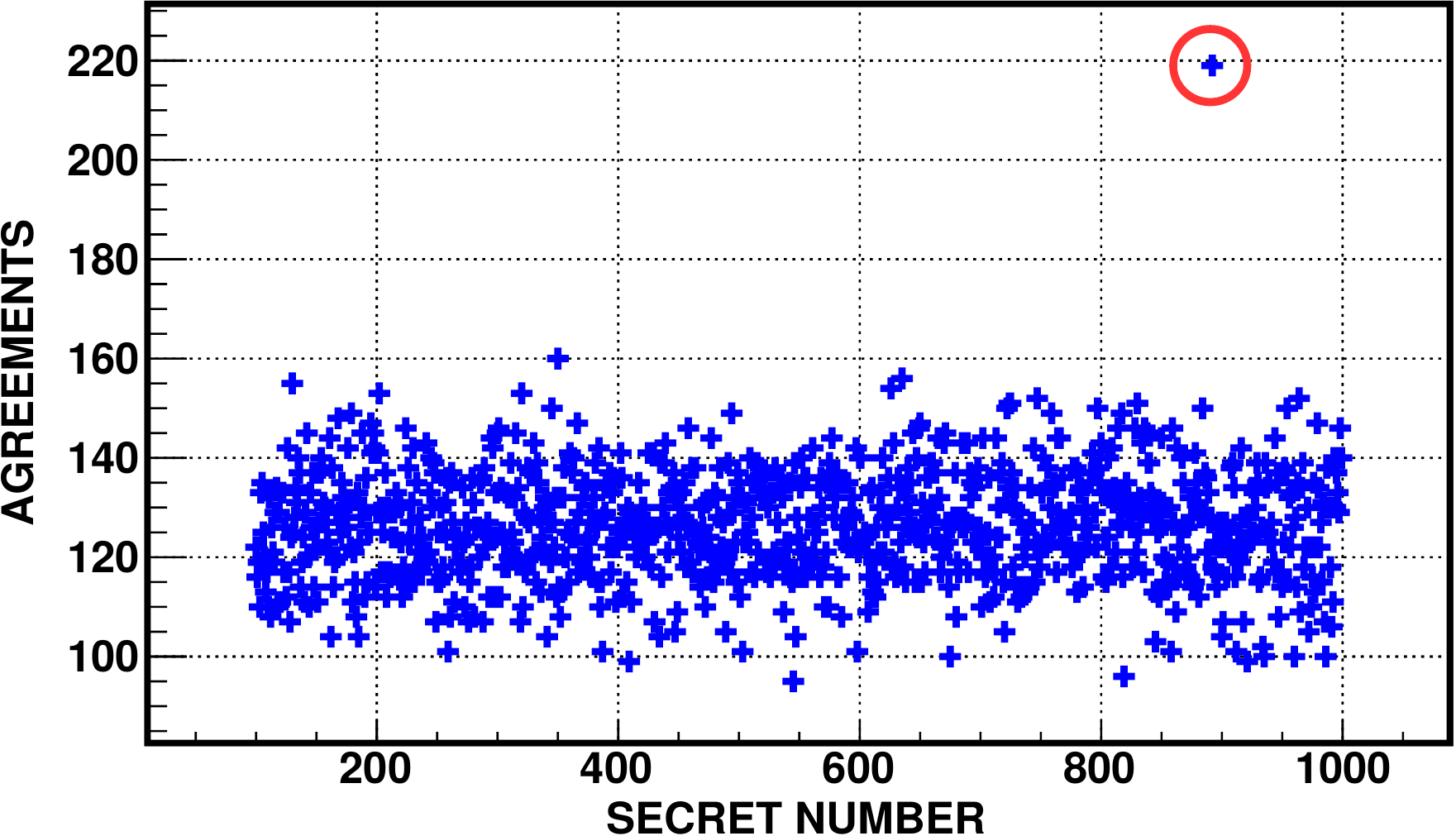}
		\caption{\label{fig:graphHash} Dependence of number of agreements on all possible three-digit secret numbers evaluated for 4 040 successfully cloned photon pairs. The revealed secret number (salt) is marked by a red circle.}	
		\end{center}
\end{figure}

To showcase the attack, we have implemented token encoding using several known hash-based functions, i.e. MD5~\cite{Rivest1992}, HMAC-SHA512, HMAC-SHA256, and HMAC-SHA1 (HMAC -- Hash-based Message Authentification Code~\cite{Bellare1996}). Typical example of encoding using SHA512 is depicted in Fig.~\ref{fig:graphHash}. In our proof-of-concept experiment, the salt has been sought only among three-digit numbers. To distinguish the secret number from noise originating from random matches, a sample of 4 040 successfully cloned photon pairs (corresponding to 101 serial numbers used in the experiment) has been evaluated. To optimise the computational resources of the attacker, the algorithm gradually refines the set of evaluated secret numbers. Periodically it removes secret numbers with low number of agreements from the list of evaluated numbers. Once the number of agreements for one secret number surpasses the average number of agreements by selected multiple of standard deviation, the algorithm ends and returns that number. Note that due to some error tolerance, the attacker does not necessarily need to recreate the original hash function. It would be enough if they found a function which error rate is below the security threshold.

The size of HMAC output of all used hash functions was set to be 40 bytes. As a consequence, the number of tokens necessary for guessing the secret number was independent on the number of digits of their serial number. For each hash function we have established how many photon pairs need to be successfully cloned in order to reveal the secret number with sufficient certainty. The results are summarised in Tab.~\ref{tab1}. The number of cloned pairs needed does not scale with the length of the salt. The salt length only increases the classical computing time. Note that these results were obtained using our experimental results where the average cloning fidelity was found to be above 80\%.

\begin{table}
\begin{ruledtabular}
\caption{Minimal number of photon pairs cloned for correct guess of the secret number (salt).}
\label{tab1}
\begin{tabular}{ll}
    hash-based function & number of pairs \\[2mm]
    \hline
 \rule{0pt}{3ex}HMAC-MD5   & 1 400 $\pm$ 16 \\
    HMAC-SHA512 & 1 192 $\pm$ 14 \\
    HMAC-SHA256 & 1 060 $\pm$ 14 \\
     HMAC-SHA1 & 1 272 $\pm$ 13 \\
\end{tabular}
\end{ruledtabular}
\end{table}

\begin{figure}
		\begin{center}
		\includegraphics[width=8cm]{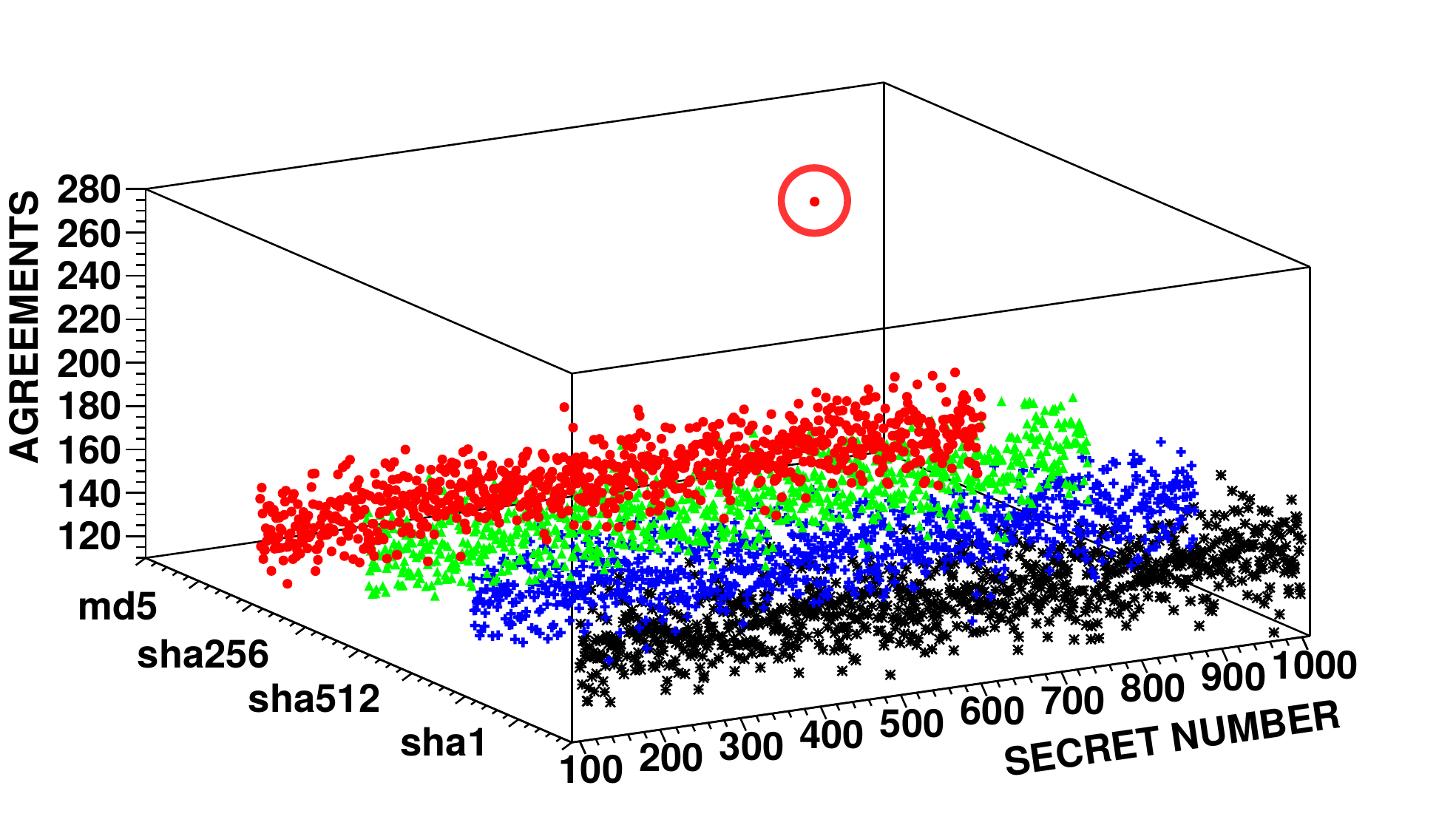}
		\caption{\label{fig:multigraphHash} Dependence of number of agreements on all three-digit secret numbers. Four different hash functions are tested. The bank used MD5 for encoding. In this plot, 4 040 successfully cloned photon pairs were analysed. The revealed secret number (salt) is marked by a red circle.}	
		\end{center}
\end{figure}

We have also performed a generalised attack in which the attacker did not know what hash function had been used for encoding. The attacker only assumes the hash function is one from a given set. In this situation, the attacker has to calculate hashes using all hash functions in this set to encode serial numbers and count numbers of agreements as described above. The plot  in Fig.~\ref{fig:multigraphHash} shows the search for the secret number among four hash functions. The tokens were encoded using MD5. Our results indicate that the correct secret number and hash function can be revealed assuming the hash function is a member of a finite set. The size of which is limited by the available time and computing power. 

\section*{Methods}
Photonic qubits were encoded as four polarisation states located on the equator of Poincaré sphere: $|D\rangle$, $|A\rangle$, $|R\rangle$ and $|L\rangle$ (i.e. diagonal linear, anti-diagonal linear, right-handed and left-handed circular polarisations). Thus, the set of possible qubit pairs~(\ref{eq:seq}) is given as
\begin{equation}
S' = \{|DR \rangle, |DL \rangle, |AR \rangle, |AL \rangle, |RD \rangle, |LD \rangle, |R  A\rangle,|LA \rangle\}\, .\label{eq:seqS}
\end{equation}

\begin{figure}[t]
		\begin{center}
		\includegraphics[scale=0.57]{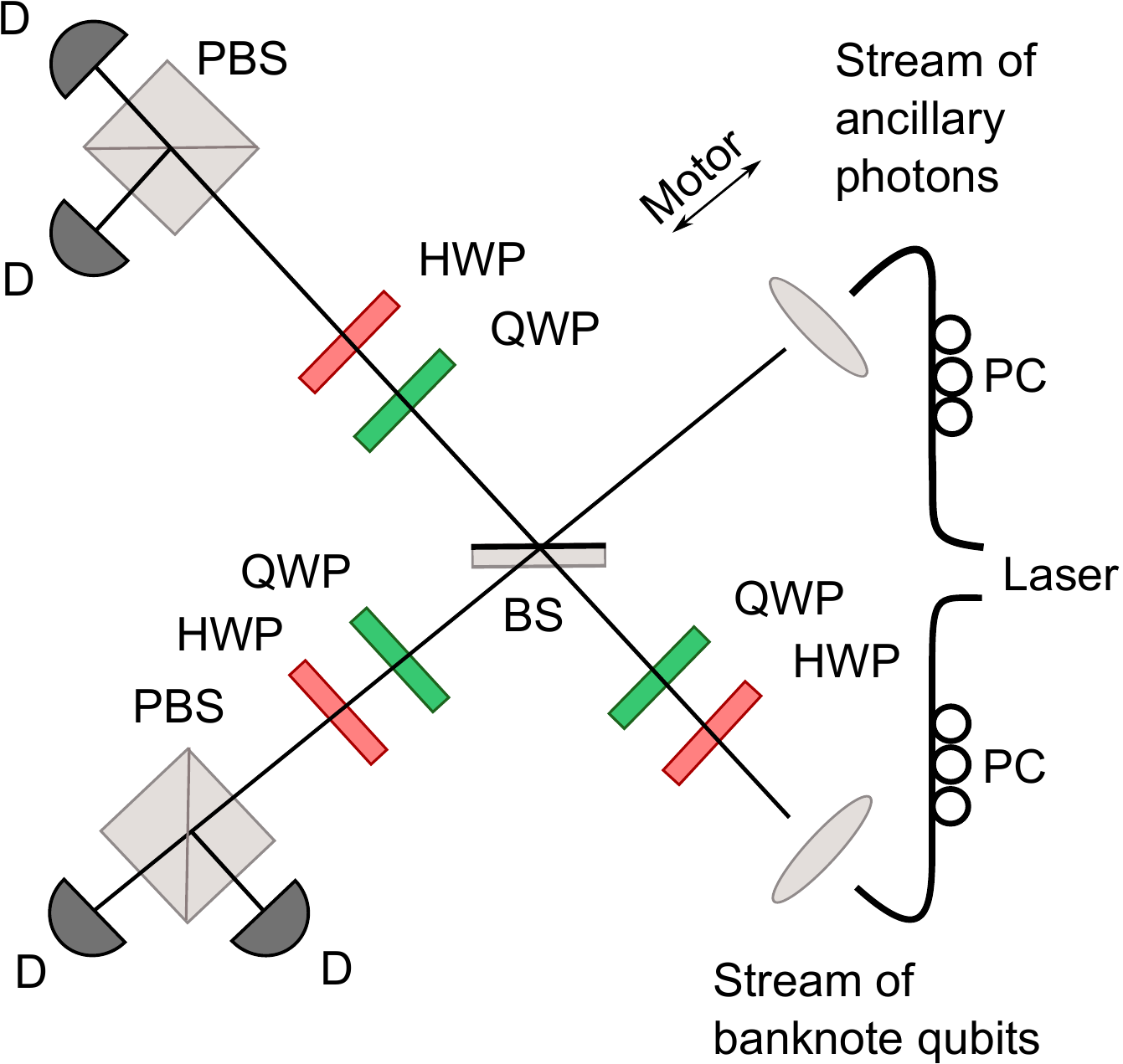}
		\caption{\label{fig:setup}Laboratory setup for the quantum sniffing experiment. The setup operates as the compromised terminal from Fig.~\ref{fig:schemeTwo}. Its components are labelled as follows: BS -- partially polarising beam splitter, QWP -- quarter-wave plate, HWP -- half-wave plate, PBS -- polarisation beam splitter, PC -- polarisation controller, D -- single-photon detector.}	
		\end{center}
\end{figure}

Experimental setup used in our experiment is shown in Fig.~\ref{fig:setup}. Photon pairs at $\lambda$ = \SI{710}{\nano\meter} are generated in a process of type-I spontaneous parametric down-conversion (SPDC) in a BBO ($\beta$-BaB$_{2}$O$_{4}$) crystal. The crystal was pumped by Paladine (Coherent) laser operating at $\lambda$ = \SI{355}{\nano\meter}. One photon from each SPDC-generated pair served as one qubit of the cloned banknote. We used a sequence of half and quarter wave plates (HWP and QWP, respectively) to implement encoding. The second photon from the SPDC-generated pair was meanwhile used as a cloning ancilla (kept horizontally polarised). Cloning is performed by an unbalanced polarisation-dependent beam splitter (BS) which implements the optimal SPCC process (for detailed theoretical description see Ref.~[\!\!\citenum{Fiurasek2003,DAriano2003,Bruss2000}], for experimental implementation see also Ref.~[\!\!\citenum{Lemr2012}]).
Subsequently, each photon is projected in the D/A or R/L measurement basis as requested by the bank (using HWPs, QWPs, and  polarisers). The process of cloning is successful only if each photon leaves BS by different output port. Therefore, we are interested in coincidences between both output arms. The detection is handled by single-photon detectors operating with detection efficiency of around 60\% and subsequent electronics. In the experiment, we have registered individual coincident detections one by one thus genuinely implementing the protocol described in the text.

\begin{figure}[t]
		\begin{center}
		\includegraphics[width=8cm]{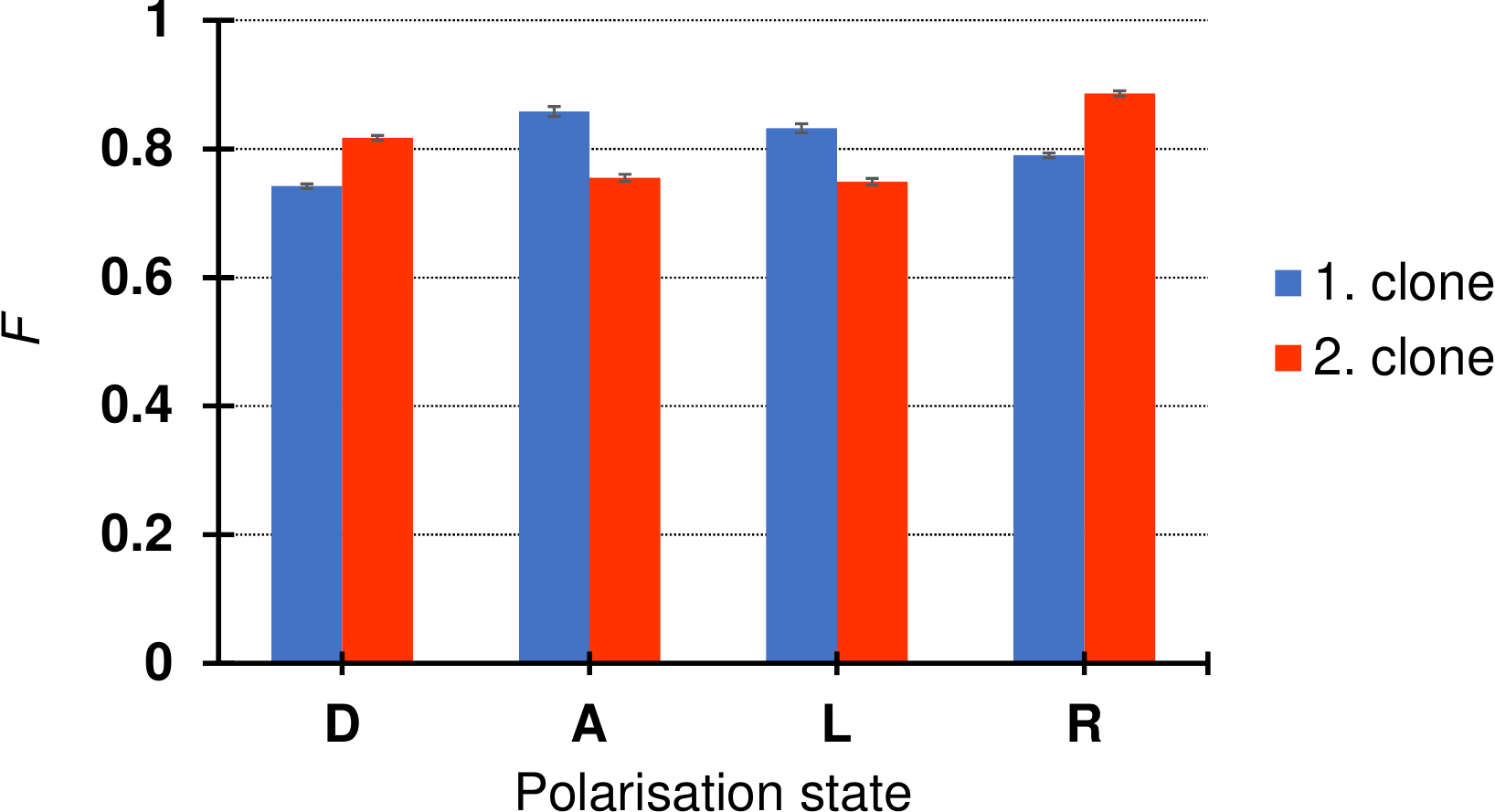}
		\caption{Average fidelity of the first and second clone of a qubit from the cloned set measured by projections in appropriate bases.}	
		\label{fig:fidel}
		\end{center}
\end{figure}

Quality of the clones was quantified by fidelity for both clones and each possible sequence qubit state (Fig.~\ref{fig:fidel}) by evaluating statistics of observed individual coincidence events. The average cloning fidelity was calculated to be (80.3 $\pm$ 0.3)\% while some clones in the two output arms had slightly different fidelities. Typical detection rate was 120 pairs per second.

\section*{Conclusion and Discussion}
We have successfully attacked a QM scheme based on QRG~\cite{Bozzio2018}. This scheme has been implemented in a form of quantum credit card containing quantum tokens.
We retrieved the secret number (salt) used for preparing quantum tokens purely by means of imperfect quantum cloning and computational analysis of measured data (see Fig.~\ref{fig:graphHash} and~\ref{fig:multigraphHash}).
By learning the exact algorithm for encoding quantum tokens, the attacker is, in principle, able to produce perfect quantum money counterfeits. It is worth noting that the optimal strategy of our attack depends mainly on a particular implementation of bank's security tolerances (e.g., losses) and chosen physical platform for implementing the attack. For instance, if the attacker uses deterministic optimal cloning even less qubit pairs is needed to perform the attack (see Fig.~\ref{fig:graphMut}).

However, the attack was feasible because the bank encoded sufficiently high number of photon pairs using the same secret number (salt) and the same hash function. From the data summarised in Tab.~\ref{tab1} we can deduce that if the bank changes, e.g., the secret number after less then 1000 photon pairs, the attacker is not able to reveal the bank's secret with sufficient certainty. This leads to further vital questions regarding tolerance of the bank to noise and threshold value losses. 

We hope that our results will stimulate further research on security of QM schemes based on QRG bringing this concept closer to becoming a fully fledged quantum technology.
Our results indicate that the correct secret number and hash function can be revealed assuming the hash function is a member of a finite set. The size of which is limited by the available time and computing power. However, this is not a fundamental limitation which might be lifted if more advanced cryptanalysis or more computing power is applied. Our results indicate that while the idea of using hash functions might be tempting, it would be ultimately more secure to store truly random sequences. The recent progress in data storage technologies may in fact enable this. Our rough estimate is that the amount of storage space required for sequences of random bits used for securing a vast amount of quantum tokens would be of similar order of magnitude as the size of Google Earth's database.

\section*{Acknowledgement}
Authors acknowledge financial support by the Czech Science Foundation under the project No. 17-10003S. The authors also acknowledge project Nos. LO1305 and CZ.02.1.01/0.0/0.0/16\_019/0000754 of the Ministry of Education, Youth and Sports of the Czech Republic and KJ also acknowledges the Palacky University internal grant No. IGA-PrF-2018-009. The authors thank CESNET for data management services.

\section*{Author contributions}
K.J., A.Č. and K.L. designed and built the experimental setup and performed the measurements. K.B. and K.L. developed the theoretical framework. The paper was written by all authors who also  participated in preparation of the manuscript and discussed the results. Figures were created by K.J. The corresponding author is K.L. (email: k.lemr@upol.cz).

\section*{Conflict of interests}

The authors declare no competing interests neither in financial matters nor in the topic of this research.


\bibliographystyle{naturemag}

\begin{thebibliography}{10}
\expandafter\ifx\csname url\endcsname\relax
  \def\url#1{\texttt{#1}}\fi
\expandafter\ifx\csname urlprefix\endcsname\relax\def\urlprefix{URL }\fi
\providecommand{\bibinfo}[2]{#2}
\providecommand{\eprint}[2][]{\url{#2}}

\bibitem{Wootters1982}
\bibinfo{author}{Wootters, W.~K.} \& \bibinfo{author}{Zurek, W.~H.}
\newblock \bibinfo{title}{A single quantum cannot be cloned}.
\newblock \emph{\bibinfo{journal}{Nature}} \textbf{\bibinfo{volume}{299}},
  \bibinfo{pages}{802--803} (\bibinfo{year}{1982}).
\newblock \urlprefix\url{http://dx.doi.org/10.1038/299802a0}.

\bibitem{Dieks1982}
\bibinfo{author}{Dieks, D.}
\newblock \bibinfo{title}{Communication by {EPR} devices}.
\newblock \emph{\bibinfo{journal}{Physics Letters A}}
  \textbf{\bibinfo{volume}{92}}, \bibinfo{pages}{271 -- 272}
  (\bibinfo{year}{1982}).
\newblock \urlprefix\url{https://doi.org/10.1016/0375-9601(82)90084-6}.

\bibitem{Wiesner1983}
\bibinfo{author}{Wiesner, S.}
\newblock \bibinfo{title}{Conjugate coding}.
\newblock \emph{\bibinfo{journal}{SIGACT News}} \textbf{\bibinfo{volume}{15}},
  \bibinfo{pages}{78--88} (\bibinfo{year}{1983}).
\newblock \urlprefix\url{http://doi.acm.org/10.1145/1008908.1008920}.

\bibitem{LemrCernoch2017}
\bibinfo{author}{{Bartkiewicz}, K.} \emph{et~al.}
\newblock \bibinfo{title}{{Experimental quantum forgery of quantum optical
  money}}.
\newblock \emph{\bibinfo{journal}{npj Quantum Information}}
  \textbf{\bibinfo{volume}{3}}, \bibinfo{pages}{7} (\bibinfo{year}{2017}).
\newblock \eprint{1604.04453}.

\bibitem{Bozzio2018}
\bibinfo{author}{Bozzio, M.} \emph{et~al.}
\newblock \bibinfo{title}{Experimental investigation of practical unforgeable
  quantum money}.
\newblock \emph{\bibinfo{journal}{npj Quantum Information}}
  \textbf{\bibinfo{volume}{4}} (\bibinfo{year}{2018}).

\bibitem{Amiri2017}
\bibinfo{author}{Amiri, R.} \& \bibinfo{author}{Arrazola, J.~M.}
\newblock \bibinfo{title}{Quantum money with nearly optimal error tolerance}.
\newblock \emph{\bibinfo{journal}{Phys. Rev. A}} \textbf{\bibinfo{volume}{95}},
  \bibinfo{pages}{062334} (\bibinfo{year}{2017}).
\newblock \urlprefix\url{https://link.aps.org/doi/10.1103/PhysRevA.95.062334}.

\bibitem{Bar-Yossef2004}
\bibinfo{author}{Bar-Yossef, Z.}, \bibinfo{author}{Jayram, T.~S.} \&
  \bibinfo{author}{Kerenidis, I.}
\newblock \bibinfo{title}{Exponential separation of quantum and classical
  one-way communication complexity}.
\newblock In \emph{\bibinfo{booktitle}{Proceedings of the Thirty-sixth Annual
  ACM Symposium on Theory of Computing}}, STOC '04, \bibinfo{pages}{128--137}
  (\bibinfo{publisher}{ACM}, \bibinfo{address}{New York, NY, USA},
  \bibinfo{year}{2004}).
\newblock \urlprefix\url{http://doi.acm.org/10.1145/1007352.1007379}.

\bibitem{Bartkiewicz2013}
\bibinfo{author}{Bartkiewicz, K.}, \bibinfo{author}{Lemr, K.},
  \bibinfo{author}{\ifmmode~\check{C}\else \v{C}\fi{}ernoch, A.},
  \bibinfo{author}{Soubusta, J.} \& \bibinfo{author}{Miranowicz, A.}
\newblock \bibinfo{title}{Experimental eavesdropping based on optimal quantum
  cloning}.
\newblock \emph{\bibinfo{journal}{Phys. Rev. Lett.}}
  \textbf{\bibinfo{volume}{110}}, \bibinfo{pages}{173601}
  (\bibinfo{year}{2013}).
\newblock
  \urlprefix\url{https://link.aps.org/doi/10.1103/PhysRevLett.110.173601}.

\bibitem{Gisin2002}
\bibinfo{author}{Gisin, N.}, \bibinfo{author}{Ribordy, G.},
  \bibinfo{author}{Tittel, W.} \& \bibinfo{author}{Zbinden, H.}
\newblock \bibinfo{title}{Quantum cryptography}.
\newblock \emph{\bibinfo{journal}{Rev. Mod. Phys.}}
  \textbf{\bibinfo{volume}{74}}, \bibinfo{pages}{145--195}
  (\bibinfo{year}{2002}).
\newblock \urlprefix\url{https://link.aps.org/doi/10.1103/RevModPhys.74.145}.

\bibitem{Bechmann1999}
\bibinfo{author}{Bechmann-Pasquinucci, H.} \& \bibinfo{author}{Gisin, N.}
\newblock \bibinfo{title}{Incoherent and coherent eavesdropping in the
  six-state protocol of quantum cryptography}.
\newblock \emph{\bibinfo{journal}{Phys. Rev. A}} \textbf{\bibinfo{volume}{59}},
  \bibinfo{pages}{4238--4248} (\bibinfo{year}{1999}).
\newblock \urlprefix\url{https://link.aps.org/doi/10.1103/PhysRevA.59.4238}.

\bibitem{Gavinsky2012}
\bibinfo{author}{Gavinsky, D.}
\newblock \bibinfo{title}{Quantum money with classical verification}.
\newblock In \emph{\bibinfo{booktitle}{2012 IEEE 27th Conference on
  Computational Complexity}}, \bibinfo{pages}{42--52} (\bibinfo{year}{2012}).

\bibitem{Pastawski2012}
\bibinfo{author}{{Pastawski}, F.}, \bibinfo{author}{{Yao}, N.~Y.},
  \bibinfo{author}{{Jiang}, L.}, \bibinfo{author}{{Lukin}, M.~D.} \&
  \bibinfo{author}{{Cirac}, J.~I.}
\newblock \bibinfo{title}{{Unforgeable noise-tolerant quantum tokens}}.
\newblock \emph{\bibinfo{journal}{PNAS}} \textbf{\bibinfo{volume}{109}},
  \bibinfo{pages}{16079--16082} (\bibinfo{year}{2012}).
\newblock \eprint{1112.5456}.

\bibitem{Georgiou2015}
\bibinfo{author}{Georgiou, M.} \& \bibinfo{author}{Kerenidis, I.}
\newblock \bibinfo{title}{New constructions for quantum money}.
\newblock In \emph{\bibinfo{booktitle}{Leibniz International Proceedings in
  Informatics, Schloss Dagstuhl Leibniz-Zentrum für Informatik}},
  \bibinfo{pages}{1--19} (\bibinfo{publisher}{Dagstuhl Publishing},
  \bibinfo{year}{2015}).

\bibitem{Wolters2017}
\bibinfo{author}{Wolters, J.} \emph{et~al.}
\newblock \bibinfo{title}{Simple atomic quantum memory suitable for
  semiconductor quantum dot single photons}.
\newblock \emph{\bibinfo{journal}{Phys. Rev. Lett.}}
  \textbf{\bibinfo{volume}{119}}, \bibinfo{pages}{060502}
  (\bibinfo{year}{2017}).
\newblock
  \urlprefix\url{https://link.aps.org/doi/10.1103/PhysRevLett.119.060502}.

\bibitem{Wang2015}
\bibinfo{author}{Wang, W.-B.}, \bibinfo{author}{Zu, C.}, \bibinfo{author}{He,
  L.}, \bibinfo{author}{Zhang, W.-G.} \& \bibinfo{author}{Duan, L.-M.}
\newblock \bibinfo{title}{Memory-built-in quantum cloning in a hybrid
  solid-state spin register}.
\newblock \emph{\bibinfo{journal}{Sci. Rep.}} \bibinfo{pages}{12203}
  (\bibinfo{year}{2015}).
\newblock \urlprefix\url{https://www.nature.com/articles/srep12203}.

\bibitem{Aaronson2012}
\bibinfo{author}{Aaronson, S.} \& \bibinfo{author}{Christiano, P.}
\newblock \bibinfo{title}{Quantum money from hidden subspaces}.
\newblock \emph{\bibinfo{journal}{Theory of Computing}}
  \textbf{\bibinfo{volume}{9}}, \bibinfo{pages}{349--401}
  (\bibinfo{year}{2013}).
\newblock \urlprefix\url{http://www.theoryofcomputing.org/articles/v009a009}.

\bibitem{Bennett1982}
\bibinfo{author}{Bennett, C.~H.}, \bibinfo{author}{Brassard, G.},
  \bibinfo{author}{Breidbard, S.} \& \bibinfo{author}{Wiesner, S.}
\newblock \bibinfo{title}{Quantum cryptography, or unforgeable subway tokens}.
\newblock In \emph{\bibinfo{booktitle}{Advances in Cryptology: Proceedings of
  CRYPTO '82}}, \bibinfo{pages}{267--275} (\bibinfo{publisher}{Plenum},
  \bibinfo{year}{1982}).

\bibitem{Bruss2000}
\bibinfo{author}{Bru\ss{}, D.}, \bibinfo{author}{Cinchetti, M.},
  \bibinfo{author}{Mauro~D'Ariano, G.} \& \bibinfo{author}{Macchiavello, C.}
\newblock \bibinfo{title}{Phase-covariant quantum cloning}.
\newblock \emph{\bibinfo{journal}{Phys. Rev. A}} \textbf{\bibinfo{volume}{62}},
  \bibinfo{pages}{012302} (\bibinfo{year}{2000}).
\newblock \urlprefix\url{https://link.aps.org/doi/10.1103/PhysRevA.62.012302}.

\bibitem{Fiurasek2003}
\bibinfo{author}{Fiur\'a\ifmmode~\check{s}\else \v{s}\fi{}ek, J.}
\newblock \bibinfo{title}{Optical implementations of the optimal
  phase-covariant quantum cloning machine}.
\newblock \emph{\bibinfo{journal}{Phys. Rev. A}} \textbf{\bibinfo{volume}{67}},
  \bibinfo{pages}{052314} (\bibinfo{year}{2003}).
\newblock \urlprefix\url{https://link.aps.org/doi/10.1103/PhysRevA.67.052314}.

\bibitem{Bartkiewicz2014}
\bibinfo{author}{Bartkiewicz, K.}, \bibinfo{author}{\ifmmode~\check{C}\else
  \v{C}\fi{}ernoch, A.}, \bibinfo{author}{Lemr, K.}, \bibinfo{author}{Soubusta,
  J.} \& \bibinfo{author}{Stobi\ifmmode~\acute{n}\else \'{n}\fi{}ska, M.}
\newblock \bibinfo{title}{Efficient amplification of photonic qubits by optimal
  quantum cloning}.
\newblock \emph{\bibinfo{journal}{Phys. Rev. A}} \textbf{\bibinfo{volume}{89}},
  \bibinfo{pages}{062322} (\bibinfo{year}{2014}).
\newblock \urlprefix\url{https://link.aps.org/doi/10.1103/PhysRevA.89.062322}.

\bibitem{Zhang2000}
\bibinfo{author}{Zhang, C.-W.}, \bibinfo{author}{Li, C.-F.} \&
  \bibinfo{author}{Guo, G.-C.}
\newblock \bibinfo{title}{Quantum clone and states estimation for n-state
  system}.
\newblock \emph{\bibinfo{journal}{Physics Letters A}}
  \textbf{\bibinfo{volume}{271}}, \bibinfo{pages}{31 -- 34}
  (\bibinfo{year}{2000}).
\newblock \urlprefix\url{https://doi.org/10.1016/S0375-9601(00)00352-2}.

\bibitem{Chefles1999}
\bibinfo{author}{Chefles, A.} \& \bibinfo{author}{Barnett, S.~M.}
\newblock \bibinfo{title}{Strategies and networks for state-dependent quantum
  cloning}.
\newblock \emph{\bibinfo{journal}{Phys. Rev. A}} \textbf{\bibinfo{volume}{60}},
  \bibinfo{pages}{136--144} (\bibinfo{year}{1999}).
\newblock \urlprefix\url{https://link.aps.org/doi/10.1103/PhysRevA.60.136}.

\bibitem{Rivest1992}
\bibinfo{author}{Rivest, R.}
\newblock \emph{\bibinfo{title}{The MD5 Message-digest Algorithm}}
  (\bibinfo{publisher}{MIT Laboratory for Computer Science},
  \bibinfo{year}{1992}).

\bibitem{Bellare1996}
\bibinfo{author}{Bellare, M.}, \bibinfo{author}{Canetti, R.} \&
  \bibinfo{author}{Krawczyk, H.}
\newblock \emph{\bibinfo{title}{Keying hash functions for message
  authentication}} (\bibinfo{publisher}{Springer-Verlag},
  \bibinfo{year}{1996}).

\bibitem{DAriano2003}
\bibinfo{author}{D'Ariano, G.~M.} \& \bibinfo{author}{Macchiavello, C.}
\newblock \bibinfo{title}{Optimal phase-covariant cloning for qubits and
  qutrits}.
\newblock \emph{\bibinfo{journal}{Phys. Rev. A}} \textbf{\bibinfo{volume}{67}},
  \bibinfo{pages}{042306} (\bibinfo{year}{2003}).
\newblock \urlprefix\url{https://link.aps.org/doi/10.1103/PhysRevA.67.042306}.

\bibitem{Lemr2012}
\bibinfo{author}{Lemr, K.}, \bibinfo{author}{Bartkiewicz, K.},
  \bibinfo{author}{\ifmmode~\check{C}\else \v{C}\fi{}ernoch, A.},
  \bibinfo{author}{Soubusta, J.} \& \bibinfo{author}{Miranowicz, A.}
\newblock \bibinfo{title}{Experimental linear-optical implementation of a
  multifunctional optimal qubit cloner}.
\newblock \emph{\bibinfo{journal}{Phys. Rev. A}} \textbf{\bibinfo{volume}{85}},
  \bibinfo{pages}{050307} (\bibinfo{year}{2012}).
\newblock \urlprefix\url{https://link.aps.org/doi/10.1103/PhysRevA.85.050307}.

\end{thebibliography}

\end{document}